\documentclass[a4paper,12pt]{article}
\usepackage{epsfig}
\usepackage{amssymb}
\usepackage{amsfonts}
\usepackage{amsmath}
\usepackage{euscript}
\usepackage{verbatim}
\usepackage{latexsym}
\usepackage{graphicx}
\usepackage{caption}
\usepackage{float}
\usepackage{subcaption}
\usepackage{wrapfig}
	\usepackage[T1]{fontenc}
	\usepackage{tikz}
	\usetikzlibrary{decorations.pathmorphing}
\usepackage{tikz}
\usetikzlibrary{shapes.geometric, arrows,patterns,snakes}
\tikzstyle{ellip} = [ellipse, minimum width=3cm, minimum height=1cm,text centered, draw=black]

\newskip\humongous \humongous=0pt plus 1000pt minus 1000pt

\newif\ifdtup

\catcode`\@=11

\@addtoreset{equation}{section}

\def\@normalsize{\@setsize\normalsize{15pt}\xiipt\@xiipt
\abovedisplayskip 14pt plus3pt minus3pt%
\belowdisplayskip \abovedisplayskip
\abovedisplayshortskip \z@ plus3pt%
\belowdisplayshortskip 7pt plus3.5pt minus0pt}

\def\small{\@setsize\small{13.6pt}\xipt\@xipt
\abovedisplayskip 13pt plus3pt minus3pt%
\belowdisplayskip \abovedisplayskip
\abovedisplayshortskip \z@ plus3pt%
\belowdisplayshortskip 7pt plus3.5pt minus0pt
\def\@listi{\parsep 4.5pt plus 2pt minus 1pt
     \itemsep \parsep
     \topsep 9pt plus 3pt minus 3pt}}

\relax

\catcode`@=12

\catcode`\@=11

\def\section{\@startsection{section}{1}{\z@}{3.5ex plus 1ex minus
   .2ex}{2.3ex plus .2ex}{\large\bf}}

\def\SymBoxes#1#2#3#4{\newdimen\un@t \un@t#3%
\raisebox{#1}{\rule{#2\un@t}{#4}\hskip-#2\un@t
\@tempdimb\un@t \advance\@tempdimb by-#4\@tempcntb#2\relax%
\@whilenum{\@tempcntb>0}\do{
\rule{#4}{\un@t}\hskip\@tempdimb \advance\@tempcntb by\m@ne}%
\hskip-#2\un@t \rule[\un@t]{#2\un@t}{#4}%
\rule[\un@t]{#4}{#4}\hskip-#4
\rule{#4}{\un@t}}\hskip-#4}                

\begin{document}

\newcommand{\beq}{\begin{equation}}
\newcommand{\eeq}{\end{equation}}
\newcommand{\bea}{\begin{eqnarray}}
\newcommand{\eea}{\end{eqnarray}}
\newcommand{\beas}{\begin{eqnarray*}}
\newcommand{\eeas}{\end{eqnarray*}}
\newcommand{\defi}{\stackrel{\rm def}{=}}
\newcommand{\non}{\nonumber}
\newcommand{\bquo}{\begin{quote}}
\newcommand{\enqu}{\end{quote}}
\renewcommand{\(}{\begin{equation}}
\renewcommand{\)}{\end{equation}}
\def \eqn#1#2{\begin{equation}#2\label{#1}\end{equation}}

\def\e{\epsilon}
\def\IZ{{\mathbb Z}}
\def\IR{{\mathbb R}}
\def\IC{{\mathbb C}}
\def\IQ{{\mathbb Q}}
\def\de{\partial}
\def\Tr{ \hbox{\rm Tr}}
\def\H{ \hbox{\rm H}}
\def\HE{ \hbox{$\rm H^{even}$}}
\def\HO{ \hbox{$\rm H^{odd}$}}
\def\K{ \hbox{\rm K}}
\def\Im{ \hbox{\rm Im}}
\def\Ker{ \hbox{\rm Ker}}
\def\const{\hbox {\rm const.}}
\def\o{\over}
\def\im{\hbox{\rm Im}}
\def\re{\hbox{\rm Re}}
\def\bra{\langle}\def\ket{\rangle}
\def\Arg{\hbox {\rm Arg}}
\def\Re{\hbox {\rm Re}}
\def\Im{\hbox {\rm Im}}
\def\exo{\hbox {\rm exp}}
\def\diag{\hbox{\rm diag}}
\def\longvert{{\rule[-2mm]{0.1mm}{7mm}}\,}
\def\a{\alpha}
\def\dag{{}^{\dagger}}
\def\tq{{\widetilde q}}
\def\p{{}^{\prime}}
\def\W{W}
\def\N{{\cal N}}
\def\hsp{,\hspace{.7cm}}

\def\br{\nonumber\\}
\def\IZ{{\mathbb Z}}
\def\IR{{\mathbb R}}
\def\IC{{\mathbb C}}
\def\IQ{{\mathbb Q}}
\def\IP{{\mathbb P}}
\def \eqn#1#2{\begin{equation}#2\label{#1}\end{equation}}

\newcommand{\C}{\ensuremath{\mathbb C}}
\newcommand{\Z}{\ensuremath{\mathbb Z}}
\newcommand{\R}{\ensuremath{\mathbb R}}
\newcommand{\rp}{\ensuremath{\mathbb {RP}}}
\newcommand{\cp}{\ensuremath{\mathbb {CP}}}
\newcommand{\vac}{\ensuremath{|0\rangle}}
\newcommand{\vact}{\ensuremath{|00\rangle}                    }
\newcommand{\oc}{\ensuremath{\overline{c}}}
\newcommand{\psiin}{\psi_{0}}
\newcommand{\phiin}{\phi_{1}}
\newcommand{\hin}{h_{0}}
\newcommand{\rh}{r_{h}}
\newcommand{\rb}{r_{b}}
\newcommand{\psibnd}{\psi_{0}^{b}}
\newcommand{\psibndp}{\psi_{1}^{b}}
\newcommand{\phibnd}{\phi_{0}^{b}}
\newcommand{\phibndp}{\phi_{1}^{b}}
\newcommand{\gbnd}{g_{0}^{b}}
\newcommand{\hbnd}{h_{0}^{b}}
\newcommand{\zh}{z_{h}}
\newcommand{\zb}{z_{b}}

\begin{titlepage}
\begin{flushright}
CHEP XXXXX
\end{flushright}
\bigskip
\def\thefootnote{\fnsymbol{footnote}}

\begin{center}
{\Large
{\bf Phases of Global AdS Black Holes \\ \vspace{0.1in} 
}
}
\end{center}

\bigskip
\begin{center}
{\large Pallab BASU$^a$\footnote{\texttt{pallabbasu@gmail.com}},  Chethan KRISHNAN$^b$\footnote{\texttt{chethan.krishnan@gmail.com}} and  P. N. Bala SUBRAMANIAN$^b$\footnote{\texttt{pnbala@cts.iisc.ernet.in}}}
\vspace{0.1in}

\end{center}

\renewcommand{\thefootnote}{\arabic{footnote}}

\begin{center}
$^a$ {International Center for Theoretical Sciences,\\
IISc Campus, Bangalore 560012, India}\\
\vspace{0.2in}

$^b$ {Center for High Energy Physics,\\
Indian Institute of Science, Bangalore 560012, India}\\

\end{center}

\noindent
\begin{center} {\bf Abstract} \end{center}

We study the phases of gravity coupled to a charged scalar and gauge field in an asymptotically Anti-de Sitter spacetime ($AdS_4$) in the grand canonical ensemble. For the conformally coupled scalar, an intricate phase diagram is charted out between the four relevant solutions:  global AdS, boson star, Reissner-Nordstrom black hole and the hairy black hole. The nature of the phase diagram undergoes qualitative changes as the charge of the scalar is changed, which we discuss. We also discuss the new features that arise in the extremal limit.

\vspace{1.6 cm}
\vfill

\end{titlepage}

\setcounter{footnote}{0}

\section{Introduction}
Understanding the phases of gravity in an asymptotically negatively curved space 
is an interesting and important conceptual problem in itself \cite{HP}. However, a new dimension opens up when you consider this problem in conjunction with holography \cite{Mal}, where  these phases get mapped to the thermal phases of quantum gauge theories \cite{witten}. 

Continuing this story, in the past decade we witnessed many interesting applications of holography in condensed matter inspired systems. Ones which are of immediate relevance to us, are a series of works \cite{Gubser,HHH2, hhh0810}, where it has been argued numerically that an Einstein-Maxwell-Scalar (EMS) system in AdS goes through a second order phase transition, in suitable parametric regimes, from a RN black hole to a hairy black hole. The resulting hairy black hole has been identified with a superconducting state (with a non-zero condensate) in the dual theory. 

In these works the background considered was the Poincar\'e patch of the AdS geometry. It is natural to ask what happens if we look at the phases of the EMS system in an asymptotically global AdS spacetime. The aim of this paper is to investigate this question in some detail. Unlike previous works in this direction (see \cite{Min1, Min2,Markeviciute:2016ivy,Bhattacharyya:2010yg} ), which considered the problem in the fixed charge\&mass ensemble, we consider this problem in the (grand) canonical ensemble\footnote{Also, we work with a conformally coupled scalar field in AdS$_4$, unlike the massless cases in AdS$_5$ considered elsewhere.}. 

Even without a scalar field, the phase structure of EMS system in global AdS is more interesting than in the Poincar\'e patch. Unlike the Poincar\'e patch case where the dual field theory lives in a flat space, here the dual theory lives on a sphere and has a mass gap coming from the curvature of the sphere. Because of this, at any finite value of the temperature $T$ and chemical potential $\mu$ there is only one phase in the Poincar\'e patch, corresponding to the RN (or Schwarzschild, if $\mu=0$) black hole. The situation changes in global AdS and we have (generalizations of) the Hawking-Page transition \cite{HP, Hawking-Reall, JME}. 

When we add a scalar, in addition to global AdS and RN black hole, we have two new hairy saddle points of gravity: one is called the boson star (see eg., \cite{Brihaye, Gentle}) and the other is the hairy black hole. Depending on the boundary chemical potential and temperature, one of the four solutions dominates the free energy landscape, giving rise to an intricate phase diagram. To have a quick idea about the phase diagram, the reader may consult Figs. \ref{q53}, \ref{ql1}. Our phase diagrams bears a rough similarity with the phase diagram of EMS in another gapped geometry, the AdS soliton \cite{HorWay2010,Takayanagi,RaamBasu}.

\section{The Setup}
\noindent The action for a Maxwell field and a charged scalar coupled to gravity \footnote{From now on we refer this system by EMS, i.e. Einstein-Maxwell-Scalar system. EM would stand for a similar system without the scalar field.}, is given by\cite{hhh0810}
\bea\label{action}
S=  -\frac{1}{16 \pi G}\int d^{4}x \sqrt{-g}\left( R + \dfrac{6}{L^{2}} -\dfrac{1}{4} F_{\mu\nu} F^{\mu\nu} -|\nabla\psi - i q A \psi|^{2}  - V(|\psi|)\right).
\eea
We will set $G=1$ in what follows. We would like to work with a time independent ansatz, which is also spherically symmetric. For the metric we pick 
\bea
ds^{2} = -g(r) h(r) dt^{2} + \dfrac{dr^{2}}{g(r)} + r^{2}\, d\Omega_{2}^{2},
\eea
and for the Maxwell and scalar fields
\bea
A = \phi(r)dt,\ \ \text{and}\ \psi = \psi(r).
\eea
The scalar field can be taken to be real, using a gauge transformation that fixes this phase \cite{hhh0810}. We also work with the potential for the scalar field of the form $V(|\psi|)= -2 M^{2}\psi(r)^{2}/L^{2}$. 
With the above ansatz, we get the equations of motion
\begin{align}
\label{eoms}
&\psi ''(r) + \frac{g'(r) \psi '(r)}{g(r)}+\frac{q^2 \psi (r) \phi (r)^2}{g(r)^2 h(r)}-\frac{V'(\psi)}{2 g(r)}+\frac{h'(r) \psi '(r)}{2 h(r)}+\frac{2 \psi '(r)}{r}= 0 ,\\
&\phi ''(r) -\frac{2 q^2 \psi (r)^2 \phi (r)}{g(r)}-\frac{h'(r) \phi '(r)}{2 h(r)}+ \frac{2 \phi '(r)}{r} = 0 ,\\
&\frac{1}{2} \psi '(r)^2 + \frac{g'(r)}{r g(r)}+\frac{q^2 \psi (r)^2 \phi (r)^2}{2 g(r)^2 h(r)}+\frac{\phi '(r)^2}{4 g(r) h(r)}- \dfrac{3}{L^{2} g(r)}-\frac{1}{r^2 g(r)}+\frac{V(\psi)}{2 g(r)}+\frac{1}{r^2} = 0 ,\\
&h'(r)-r h(r) \psi '(r)^2 -\frac{q^2 r \psi (r)^2 \phi (r)^2}{g(r)^2} = 0.
\end{align}
What we have to look for are solutions, with the asymptotic behaviour of global AdS. The set of equations have the following scaling symmetries for the functions
\begin{itemize}
\item $r \rightarrow a r, \ \ q\rightarrow \frac{q}{a}, \ {\rm and } \ L\rightarrow a L $. With this rescaling, one could set $L =1$.
\item $h \rightarrow \bar{h}= a^{2} h, \ \phi \rightarrow \bar{\phi} =  a \phi, \ {\rm and }\ t \rightarrow \bar{t}= \frac{t}{a}$, so that the time part of the metric becomes
\bea
-g h dt^{2} = - g \dfrac{\bar{h}}{a^{2}}dt^{2} = -g \bar{h} d\bar{t}^{2}.
\eea
Now, since we need the space to be asymptotically global AdS, we need to have\footnote{We will suppress the \emph{bar} on $\bar{h}$ in what follows with the understanding that we are always using the rescaled metric.}
\bea
\lim_{r\rightarrow\infty}\bar{h}(r) = 1,\ \text{i.e} \ \ \bar{h}(\rb)= a^{2} h(\rb)=1 \Rightarrow a = \dfrac{1}{\sqrt{h(r\rightarrow\infty)}}.
\eea
\end{itemize} 
We will be looking at conformally coupled scalar, which sets $M =1$. The asymptotic expansion for such a scalar for any asymptotically-$AdS_{4}$ space has the falloff
\bea
\lim_{r\rightarrow\infty} \psi(r) \simeq \dfrac{\psi_{(1)}}{r} + \dfrac{\psi_{(2)}}{r^{2}} + \dots \; .
\eea
We will consider the boundary conditions for the systems at $r\rightarrow\infty$ of the following form,
\bea
g(r\rightarrow\infty) \simeq 1 +r^{2}, \ h(r\rightarrow\infty)= 1,\ \psi_{(1)}= 0,\ \text{and} \  \phi(r\rightarrow\infty) = 2 \mu.
\eea
(See footnote 3 for a discussion on the origin of the factor of 2 in the definition of $\mu$.)
Depending on whether we are looking for solutions with or without the horizon, we have to specify the boundary condition at $r=0$, if there is no horizon, or at $r=\rh$ which is the location of the horizon. This boundary condition will be discussed separately, depending on the solutions that we will be looking at.\\
We use the scaling symmetry to set $L=1$, for all the numerical solutions in the following.

\section{The different solutions}
As we will see in the following discussion in detail, there exist four different (classes of) solutions for the action in \eqref{action}. Of them, global AdS and RNAdS are solutions where the scalar is zero all throughout and will be discussed first. For non-trivial scalar profiles the solutions are called boson star and the hairy black hole, which will be discussed in more detail. The thermodynamic favourability of these four solutions are based on the free energy of the respective solution. 
\subsection{Global AdS}
The vacuum solution of the action in \eqref{action}, is the global AdS, given by
\bea
g(r) = 1+ \dfrac{r^{2}}{L^{2}}, \ h(r) =1, \ \phi'(r) =0,\ \text{and} \ \psi(r) = 0, 
\eea
We can choose $\phi(r)=2 \mu$, a constant. This solution exists for any chemical potential and temperature, but it is not necessarily the global minimum of the free energy. At $T=\mu=0$ the only possible solution of \eqref{action} is global AdS.
 
\subsection{RNAdS}
\begin{figure}[H]
\begin{center}
\includegraphics[width=.5\textwidth]{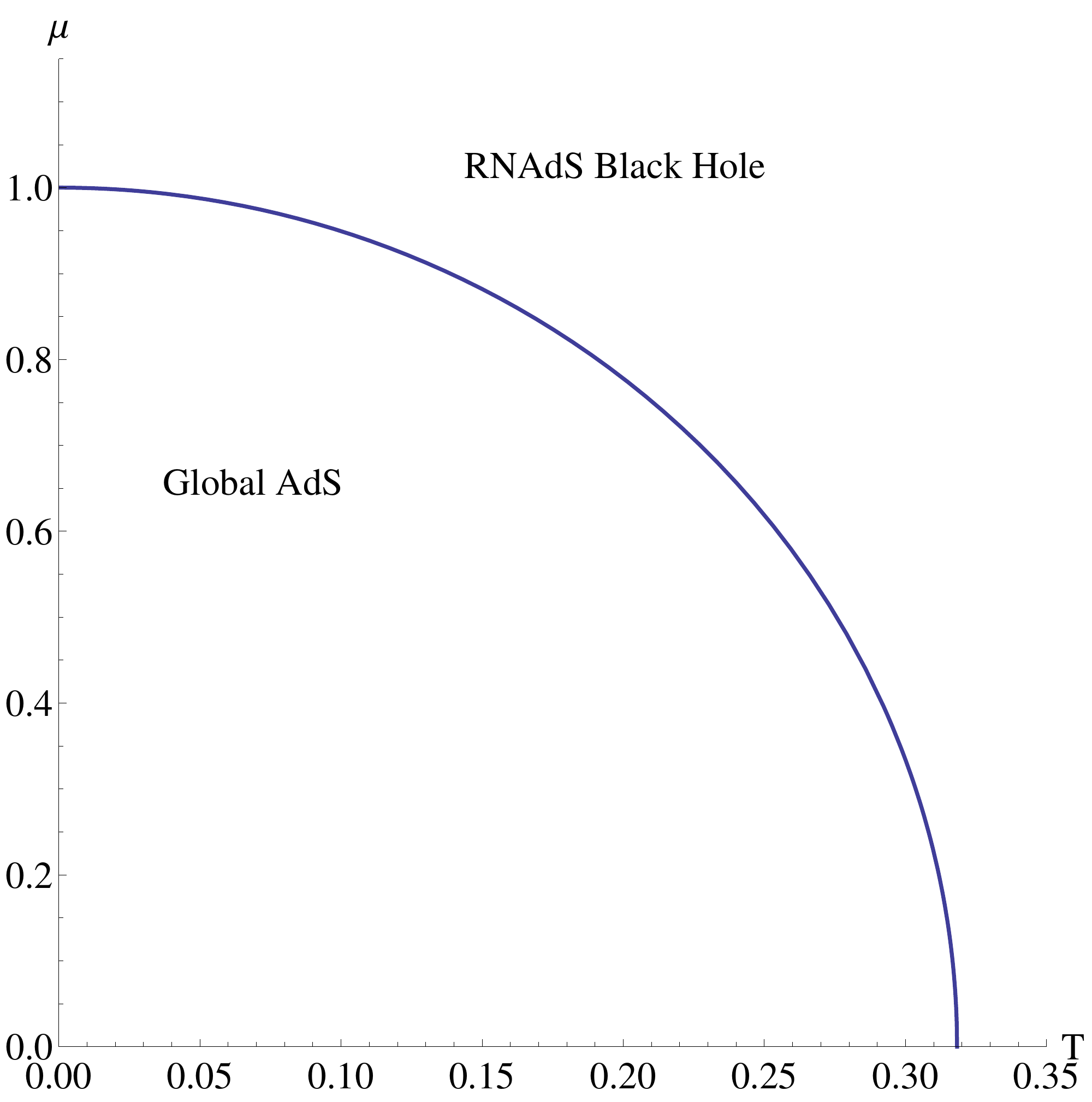}
\end{center}
\caption{Phase diagram for RNAdS black hole}
\label{rnadsphasedig} 
\end{figure}

With the scalar in \eqref{action} turned off, we impose the boundary conditions at the horizon for the metric
\bea
g(\rh)=0,\ \text{and} \ \phi(\rh) =0,
\eea
which is required to ensure the regularity of the gauge connection. This is enough to completely fix the solution, which is given by
\bea
g(r) = 1 + \dfrac{\rho^2}{4 r^2} +\dfrac{r^2}{L^2}- \dfrac{L^2 \rho^2 + 4 L^2 \rh^2 + 4 \rh^4}{4 L^2 \rh \, r}, \, h(r) =1, \, \phi(r) =\rho\left(\dfrac{1}{\rh} - \dfrac{1}{r}\right).
\eea
The number of scaling symmetries in the case of global AdS is one less than that in the Poincar\'e patch\cite{hhh0810}. This means that one cannot set the horizon and $L$ to be 1 at the same time. This means that for different horizon radii, the solutions are different, for a given value of $L$. The temperature for the metric ansatz that we have, by imposing that there is no conical singularity after Wick rotation, is given by 
\bea
T = \dfrac{1}{4\pi} \dfrac{g'(\rh)}{\sqrt{h(r\rightarrow \infty)}}.
\eea
For the RNAdS system, we get the temperature and chemical potential to be
\bea 
T =\frac{1}{4 \pi}\left( \dfrac{1}{\rh} - \dfrac{\rho^2}{4 \rh^3} + \dfrac{3 \rh}{L^2}\right),\ \text{and} \ \mu = \dfrac{\rho}{2 \rh}.
\eea
The free energy of the system can be computed by evaluating the classical action for the solution ($I=\beta F$), and subtracting out the same of global AdS \cite{witten}, for details see Appendix A. It can alternately be evaluated using the ADM mass and charge of the system, as
\bea\label{therm}
F = E -T \, S -\mu \, Q,
\eea
where\footnote{We are following the conventions of \cite{JME}, except for the fact that we have an extra factor of $1/4$ in the Maxwell piece in the action as in \cite{HorWay2010}. This is the source of the extra factor of 2 between $\rho$ and $Q$, as well as between $\phi(r\rightarrow \infty)$ and $\mu$.} $Q=\rho/2$. 
Both approaches lead to the free energy
\bea
F = -\dfrac{4 \rh^4 + L^2 (\rho^2 - 4 \rh^2)}{16 \rh}
\eea
for the RN black hole. From the free energy one may chart down the phase diagram. The RNAdS black hole becomes the dominant phase when the free energy goes negative. Imposing this condition, we can solve for $T$ in terms of $\mu$ or vice versa, which demarcates the two phases. These two phases are separated by a line of first order transition know as Hawking-Page transition \cite{HP}(see Fig.\ref{rnadsphasedig}). 

\subsection{The different instabilities and hairy solutions}
We will now look into the hairy solutions, by turning on the scalar field and finding the instability of global AdS and RNAdS for forming scalar hair. The corresponding solutions are boson star and hairy black hole respectively. Also, we will look into the properties of the hairy solutions, and investigate the phase structure of the full system.
\subsubsection{Boson star instability}
As the chemical potential increases to a critical value, say $\mu =\mu_{c 1}$, the scalar field will develop a zero mode, i.e. we will have  $\psi_{(1)}=0$ (scalar condensate), without the formation of a horizon. This configuration is called Boson star \cite{Brihaye}. Here, boundary conditions at $r=0$ is given by $\phi'(0)=0,h'(0)=0,\psi'(0)=0,g'(0)=0$ which ensure that there is no kink at $r=0$. Also, we set $g(0)=1$ and $h(0)$ is set to an arbitrary value, as we will rescale it to make sure that the asymptotic boundary conditions are satisfied. \\

To understand the onset of the formation of the scalar zero mode, we look at the equation of motion of scalar in global AdS. A probe computation is enough to specify the instability because for very small scalar profile, the back-reaction is negligible. By a probe computation here, we mean taking $\psi(r)\rightarrow \epsilon \psi(r)$ with $\epsilon\ll 1$. Now, looking only upto linear order terms in $\epsilon$, the two equations coming from the Einstein equation and the Maxwell equation decouple from the scalar, giving the RNAdS solution. The scalar equation of motion then becomes a homogeneous equation in this background, and is given by
\bea
\psi ''(r) + \left(\frac{2 r}{L^2 \left(\frac{r^2}{L^2}+1\right)}+\frac{2}{r}\right) \psi '(r) + \left(\frac{2 M^2}{L^2 \left(\frac{r^2}{L^2}+1\right)}+\frac{4\mu ^2 q^2}{\left(\frac{r^2}{L^2}+1\right)^2}\right) \psi (r) = 0.
\eea
 The solution of this equation with the above mentioned boundary conditions can be analytically determined and is given by
 \bea
\psi(r) = C \ \dfrac{\sin(2 q\mu \tan^{-1}r)}{r}, \ \text{where} \ q\mu =n, \ n\in \mathbf{Z},
\eea
where $C$ is arbitrary, and signifies the overall scaling freedom we get in the probe limit. The quantization in $n$  happens because AdS is like a box, as far as the scalar is considered. 

We will look at the first non trivial solution, which is $n=1$, for the boson star phase, as there are no nodes in the scalar profile. For the $n=1$ mode, we have the relation $\mu \, q =1$, which gives the chemical potential ($\mu_{c1}$) at which the global AdS phase goes to the boson star phase, for a given $q$. It can be seen that for large $q$ the condensate can be formed by a very small $\mu$ and vice-versa.
 
\begin{figure}
    \centering
    \begin{subfigure}[b]{0.4\textwidth}
        \includegraphics[width=\textwidth]{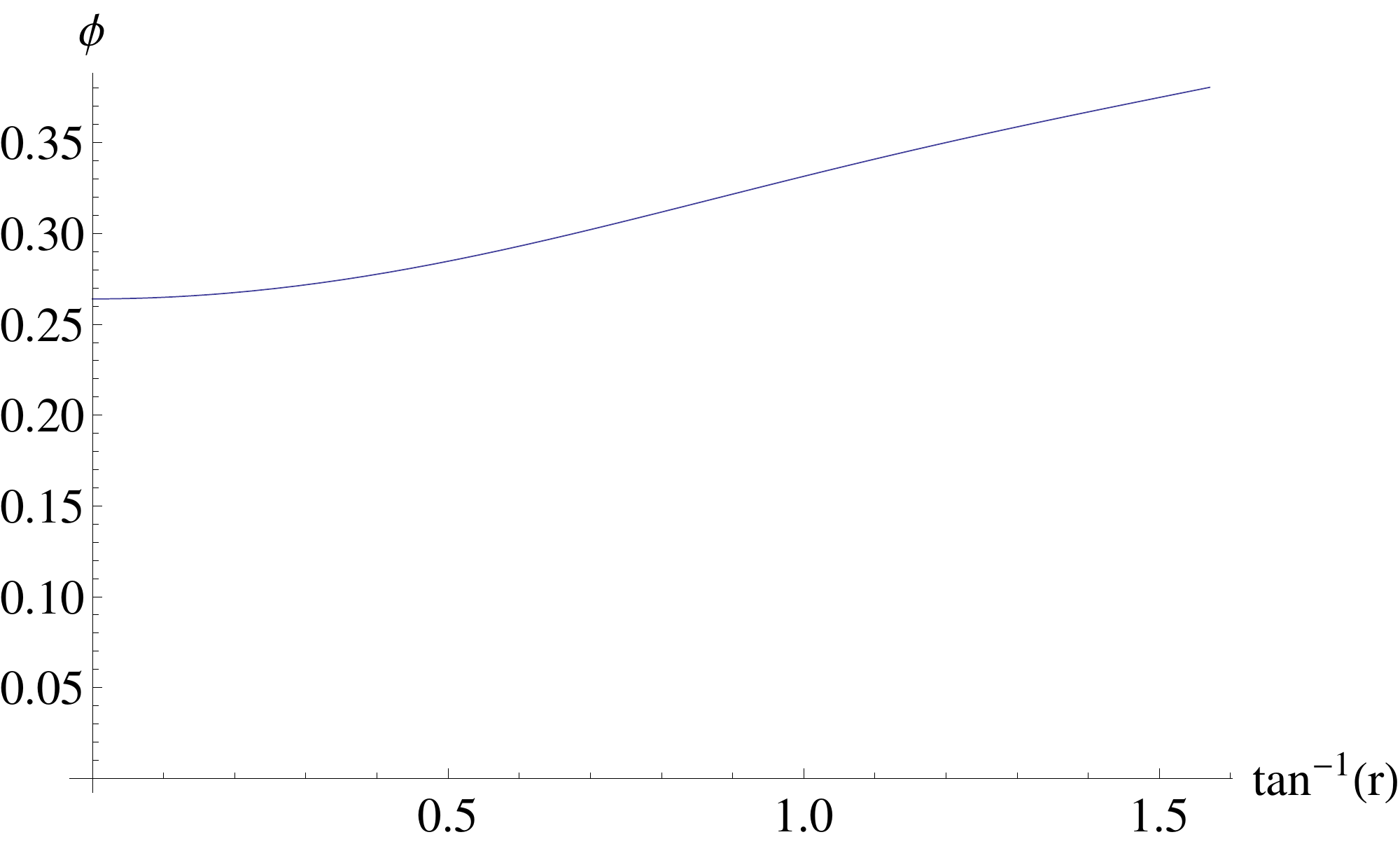}
        \caption{}
     \end{subfigure}
    \begin{subfigure}[b]{0.4\textwidth}
        \includegraphics[width=\textwidth]{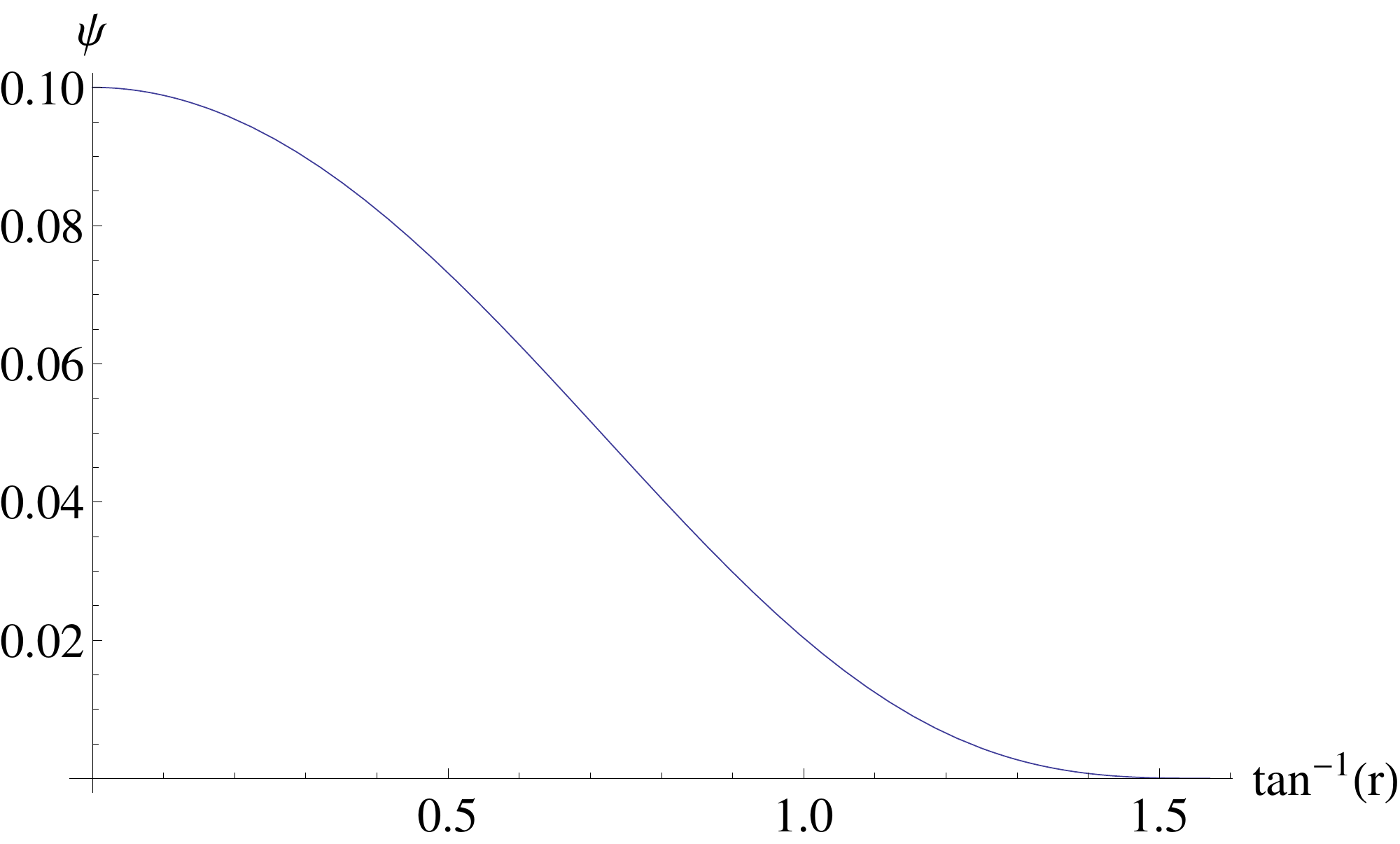}
        \caption{}
        \end{subfigure}
     \caption{The profiles of $\psi(r)$ and $\phi(r)$ in the fully backreacted solution for boson star.}\label{bsprofiles}
\end{figure}

As we move away from the probe case, the solutions can be found numerically\footnote{Except for determining the point of onset of the instability, we do not use the probe computation anywhere.}. Here we have plotted the fully back reacted solution for $q=10$ and $2 \mu=0.3804$, see Fig.(\ref{bsprofiles}). Since it is the fully back-reacted solution, the relation $q\mu =1$ doesn't hold.
 
\subsubsection{Hairy Black Hole instability}
A similar analysis can be done in a RNAdS background, for a probe scalar for the formation of a hair. The scalar equation of motion is solved in the RNAdS background, with the boundary conditions at the horizon $r =\rh $ for the scalar given by $\psi(\rh )= \psi_{0}$, and the first derivative $\psi'(\rh )$ fixed by the consistency of the series expansion of the equations of motion around the horizon. For the asymptotic boundary, $\psi_{(1)} = 0$. The scalar equation of motion in the RNAdS background is given by,
\bea
\psi ''(r) +\left(\frac{\rho^2 (r-2\rh)+4 r \rh \left(2 r^3+\rh^3+\rh\right)}{r (r-\rh) \left(4 r \rh \left(r^2+r \rh+\rh^2+1\right)-\rho^2\right)}+\frac{2}{r}\right) \psi '(r) \qquad\qquad \nonumber\\ \qquad \qquad +\left(\frac{8r^{2} \left[2q^{2} \rho^{2} - M^{2}\rh (\rho^{2} - 4r\rh (1+r^{2}+ r \rh +\rh^{2}))\right]}{(r-\rh )(\rho^{2} - 4r\rh (1+r^{2}+ r \rh +\rh^{2}))^{2}} \right) \psi(r) = 0.
\eea
For a given value of $q$ and $\rh $, we take an arbitrary value for $\psi_{0} $ (because the probe equation is homogeneous, the solutions are rescalable) and find 
$\rho$ such that the scalar develops a zero mode $\psi_{(1)}=0$, by a numerical shooting method. This fully determines the RNAdS black hole background as we know $\rh$ and $\mu=\rho/2 \rh$, from which we can also find the temperature $T$. Repeating this for different values of $\rh$ keeping $q$ fixed will give the instability curve\footnote{Where this line features will be discussed while considering the full phase diagram}. This instability indicates a second order phase transition from the RNAdS to hairy black hole phase. For a fixed $q$, and varying $\rh $ we get a curve in the $\mu-T$ plane, which demarcates the two phases. As in the case of the boson star, here too, the larger the value of $q$, the lower the chemical potential $\mu$ required to form a condensate. 

\begin{figure}
    \centering
    \begin{subfigure}[b]{0.4\textwidth}
        \includegraphics[width=\textwidth]{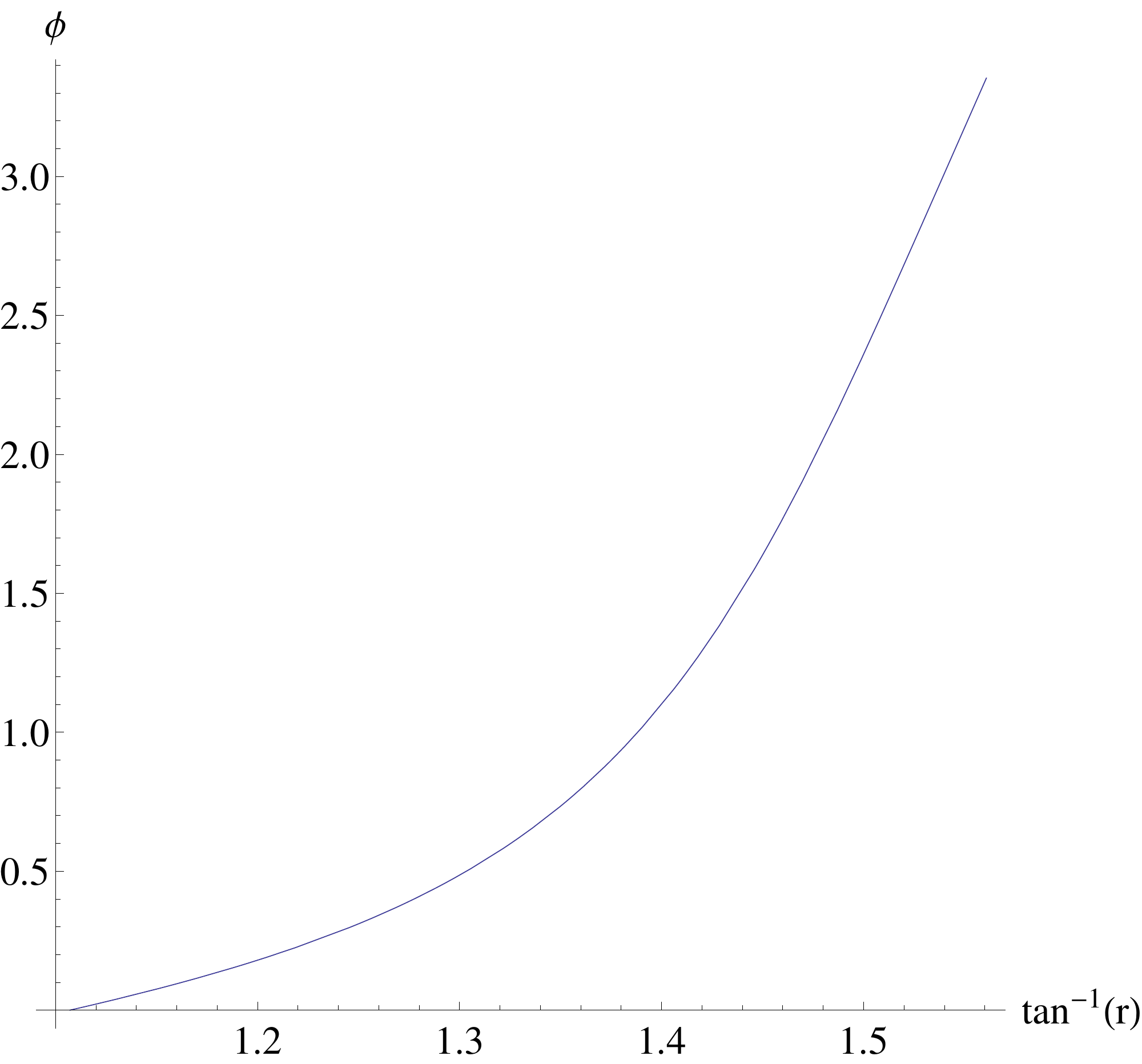}
        \caption{}
         \end{subfigure}
    \begin{subfigure}[b]{0.4\textwidth}
        \includegraphics[width=\textwidth]{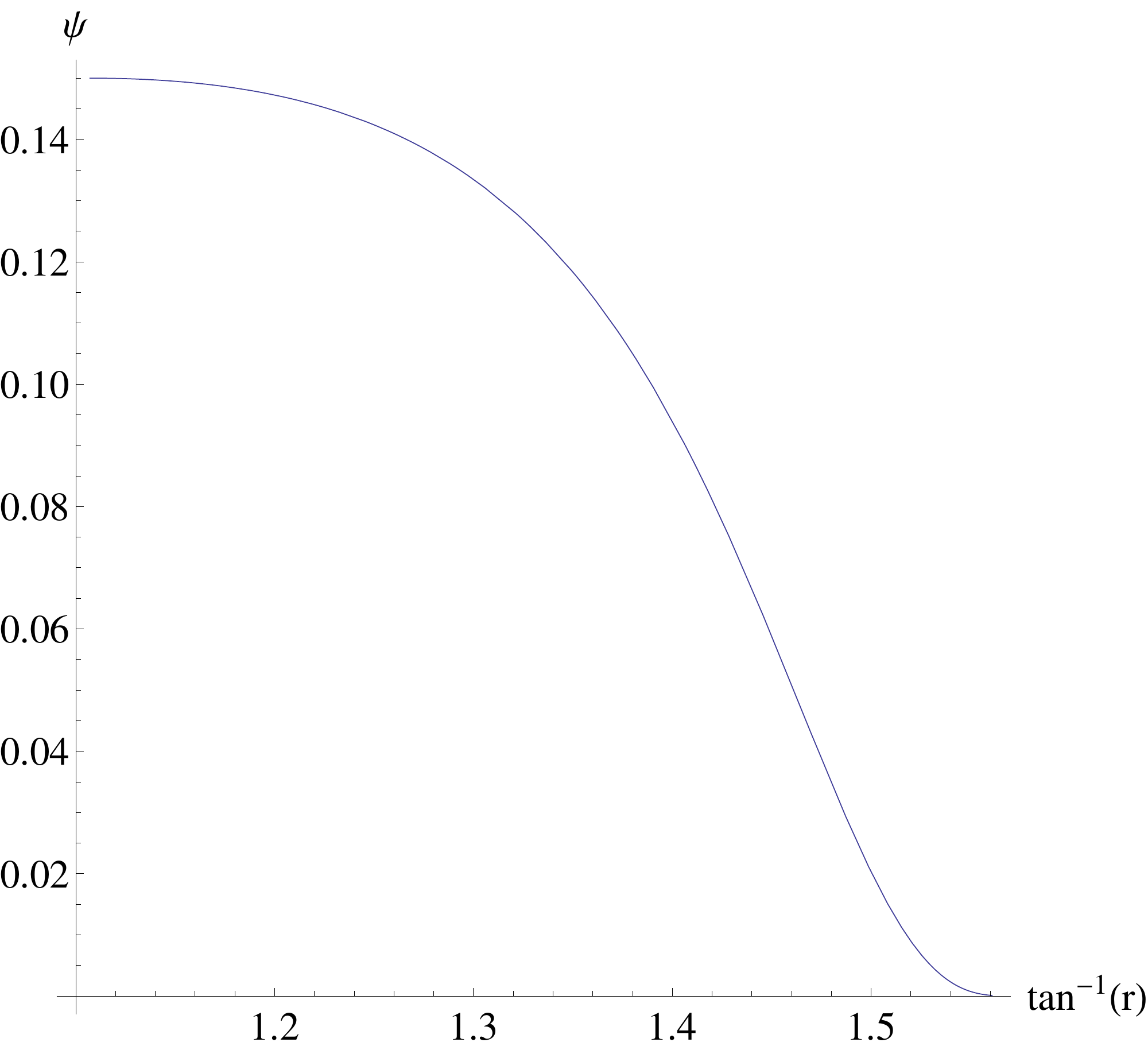}
        \caption{}
         \end{subfigure}
     \caption{Sample profiles of $\psi(r)$ and $\phi(r)$ in the fully backreacted solution for hairy black hole.}\label{bhprofiles}
\end{figure}

We also can construct the fully back-reacted hairy solution numerically, see Fig.(\ref{bhprofiles}).

%
%
%

\section{Phase diagram}
The phase diagram of the Einstein-Maxwell-Scalar (EMS) system is more complicated than the EM system. In EM system there are only two possible radially symmetric solutions\cite{Hawking-Reall}, i.e. global AdS and a charged BH. Whereas, as we discussed, EMS system has two more types of solutions, which are boson stars and hairy black holes. Depending on the temperature and the chemical potential, one of the four possible solutions becomes the dominant phase (i.e. the phase with least free energy) of the theory. The intricateness of the phase diagram depends on the value of $q$.

As we saw in the boson star instability discussion, the chemical potential that sets up this instability is given by $\frac{1}{q}$. In Fig.\ref{rnadsphasedig}, we can see that $\mu =1$ for $T=0$, above which the global AdS is not the stable solution. Since the boson star is a second order transition from global AdS, it does not exist once the chemical potential required for this instability is not within the global AdS phase, i.e. if $\mu>1$, or in other words, when $q<1$. Hence, there will be a qualitative difference in the phase diagram for $q>1$ and $q<1$, which we will deal with separately.

\subsection*{$ \infty >q>1$}
\begin{figure}[H]
\begin{center}
\begin{tikzpicture}
\coordinate (a) at (0,0) {};
\coordinate (b) at (8,0) {};
\coordinate (c) at (0,8) {};
\coordinate (d) at (0,2.5) {};
\coordinate (e) at (4.5,0) {};
\coordinate (f) at (4,2.5) {};
\coordinate (g) at (3,4) {};
\coordinate (h) at (2.5,7.5) {};
\coordinate (i) at (7,7) {};

\draw[ultra thick] (a) -- (b);
\draw[ultra thick] (a) -- (c);
\draw[ultra thick,dashed] (d) --node[midway, below right] {S-1} (f);
\draw[blue,ultra thick] (e) to [out=90,in=300] node[midway, below right] {F-1} (f);
\draw[green,ultra thick] (f) to [out=120,in=280] node[midway, below right] {  \;\;F-2} (g);
\draw[purple,ultra thick] (g) to [out=100,in=280] node[midway, below right] {F-3}(h);
\draw[red,ultra thick,dashed] (g) --node[midway, below right] {S-2} (i);
\node at (7.5,-0.5) {$T$};
\node at (-0.5,7.5) {$\mu$};
\node at (6,2.5) {$\framebox{RNAdS}$};
\node at (1.5,1) {$\framebox{global AdS}$};
\node at (1.2,6) {$\framebox{Boson Star}$};
\node at (5,7) {$\framebox{Hairy BH}$};

\end{tikzpicture}
\end{center}
\caption{Schematic phase diagram for $\infty  >q >1$.}
\label{fig:phase1}
\end{figure}
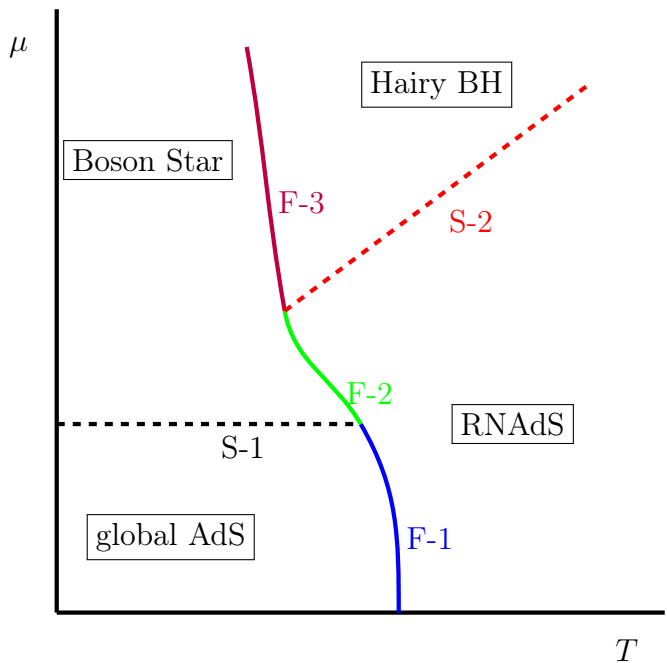

In the case of $q>1$, the boson star instability happens at $\mu<1$. The schematic phase diagram is given in Fig.\ref{fig:phase1}. These phase diagrams are found for a fixed $q$, by a mixture of analytic and numerical methods where appropriate.

For sufficiently small $\mu$, which is less than required for the boson star instability ($\mu_{c1}$), we will have global AdS and RNAdS as the phases in the theory, demarcated by a first order phase transition, which is the Hawking-Page transition, given by the curve \emph{F-1} in the figure. As we increase $\mu$, beyond $\mu_{c1}$, there is a second order phase transition \emph{S-1}, from global AdS to boson star. These two phase boundaries are analytically tractable, as we saw. The Hairy black hole instability happens at a larger value of $\mu$, say $\mu_{c2}$, for any given temperature. The two phases in the region $\mu_{c1}<\mu<\mu_{c2}$, will be separated by a first order transition, given by the curve \emph{F-2}. We can find the curve \emph{F-2}, semi-analytically, as follows. First we numerically find the fully back-reacted boson star solutions, for some value of $\psi(0)$, above the line \emph{S-1}. A given value of $\psi(0)$ corresponds to a unique boson star solution, so that fixes its $\mu$ and $F$. It is now straightforward to analytically compute the temperature $T$ of the RNAdS black hole with this same $\mu$ and $F$ and that gives us a point on the \emph{F-2} curve in the $T-\mu$ plane. Changing the value of $\psi(0)$ and repeating the process produces the full \emph{F-2}.

As we keep increasing the value of $\mu$ further, the RNAdS black hole is unstable towards formation of scalar hair. The RNAdS and hairy black hole phases are separated by a second order phase transition, given by the curve \emph{S-2}. This curve is found numerically, by looking at the hairy black hole instability, as discussed in the previous section, for different $\rh$. Now, the hairy black hole and boson star phases will also have a first order phase boundary given by the curve \emph{F-3}, which is again found numerically as we discuss presently.

The way in which the phase diagram is computed for the global AdS is slightly different from the way in which it is done for the case of the Poincar\'e patch\cite{HorWay2010}. For the Poincar\'e patch, since there is an additional coordinate rescaling freedom, what one does is take a value of $\mu$, rescale all the solutions to have the same value of $\mu$, and then evaluate the free energies for different $T$, thus finding the phase boundaries. The lack of that rescaling in global AdS does not become an issue while evaluating the curves \emph{F-1} and \emph{F-2} because they are (semi-)analytic. Determining \emph{S-1,S-2} is also straightforward because they are probe computations as we explained, which rely on the fact the phases with condensates have lesser free energy than phases with condensate.

The curve \emph{F-3} is more difficult to find, precisely because of the missing coordinate rescaling freedom, and the fact that both the phases to be compared are fully back-reacted solutions, hence, completely numerical. What we do first is to find the free energy ($F$) of the boson star for different values of $\mu$, and use a numerical fit to get $F$ as a function of $\mu$. One useful fact to keep in mind for the hairy black hole solutions is that for a fixed $\rh $, as we increase the value of $\psi_{0}$, the chemical potential and the temperature both increase, the latter very slowly. Thus, to obtain the curve \emph{F-3}, we take horizon radius slightly less than that which corresponds to the black hole 
at the intersection of the curves \emph{F-2} and \emph{S-1} (we will call this critical horizon radius, $r_{h22}$), and keep increasing $\psi_{0}$, simultaneously evaluating $\mu$, $T$ and the free energy. The curve \emph{F-3} is gotten by finding the $(T,\mu)$ values for which the free energies of the hairy black hole and the boson star match, by repeating the process mentioned above for smaller and smaller values of $\rh$ compared to $r_{h22}$.

\begin{figure}[H]
    \centering
    \begin{subfigure}[b]{0.45\textwidth}
        \includegraphics[width=\textwidth]{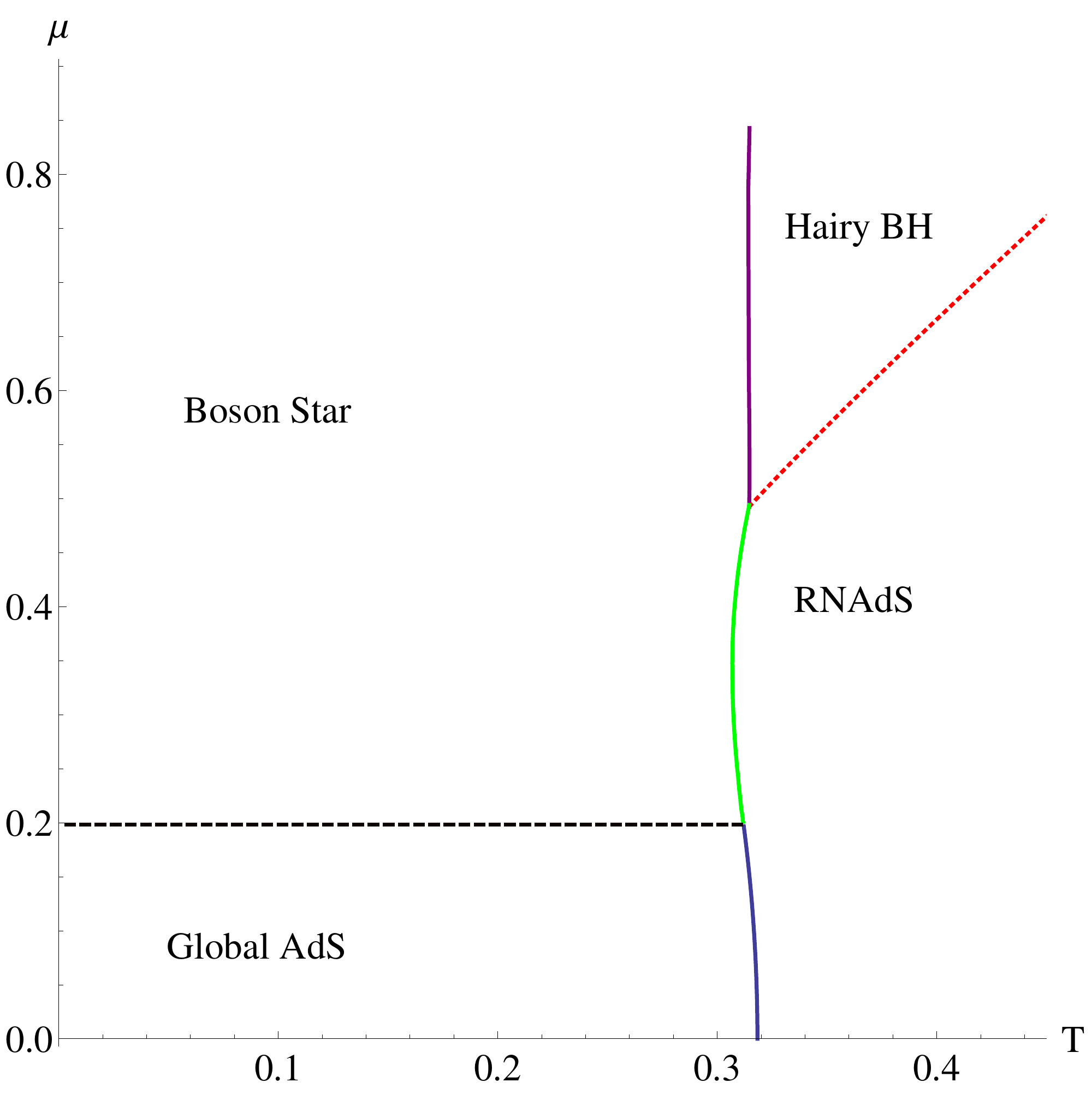}
        \caption{}
           \end{subfigure}
    \begin{subfigure}[b]{0.45\textwidth}
        \includegraphics[width=\textwidth]{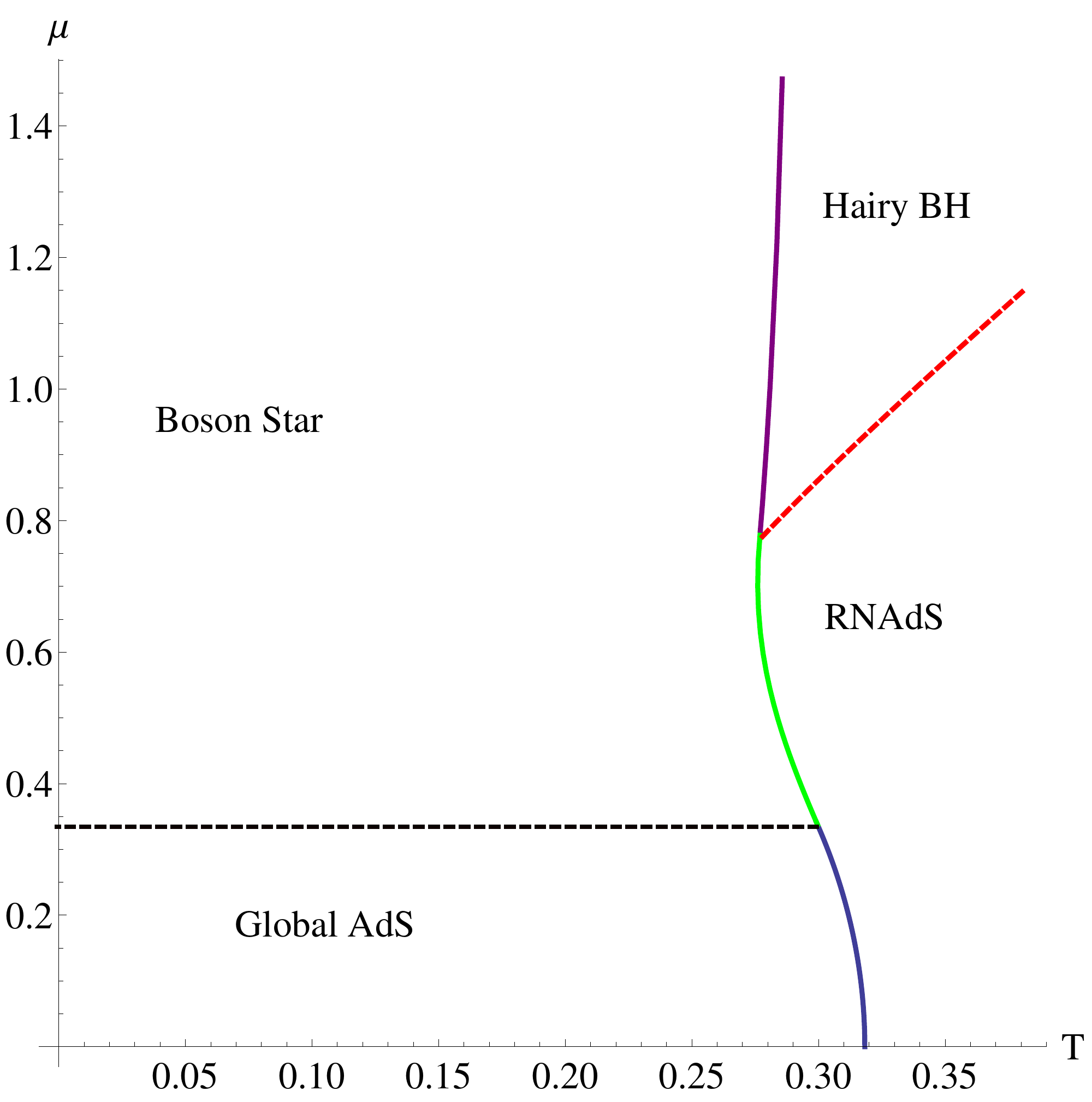}
        \caption{}
            \end{subfigure}
\begin{subfigure}[b]{0.45\textwidth}
        \includegraphics[width=\textwidth]{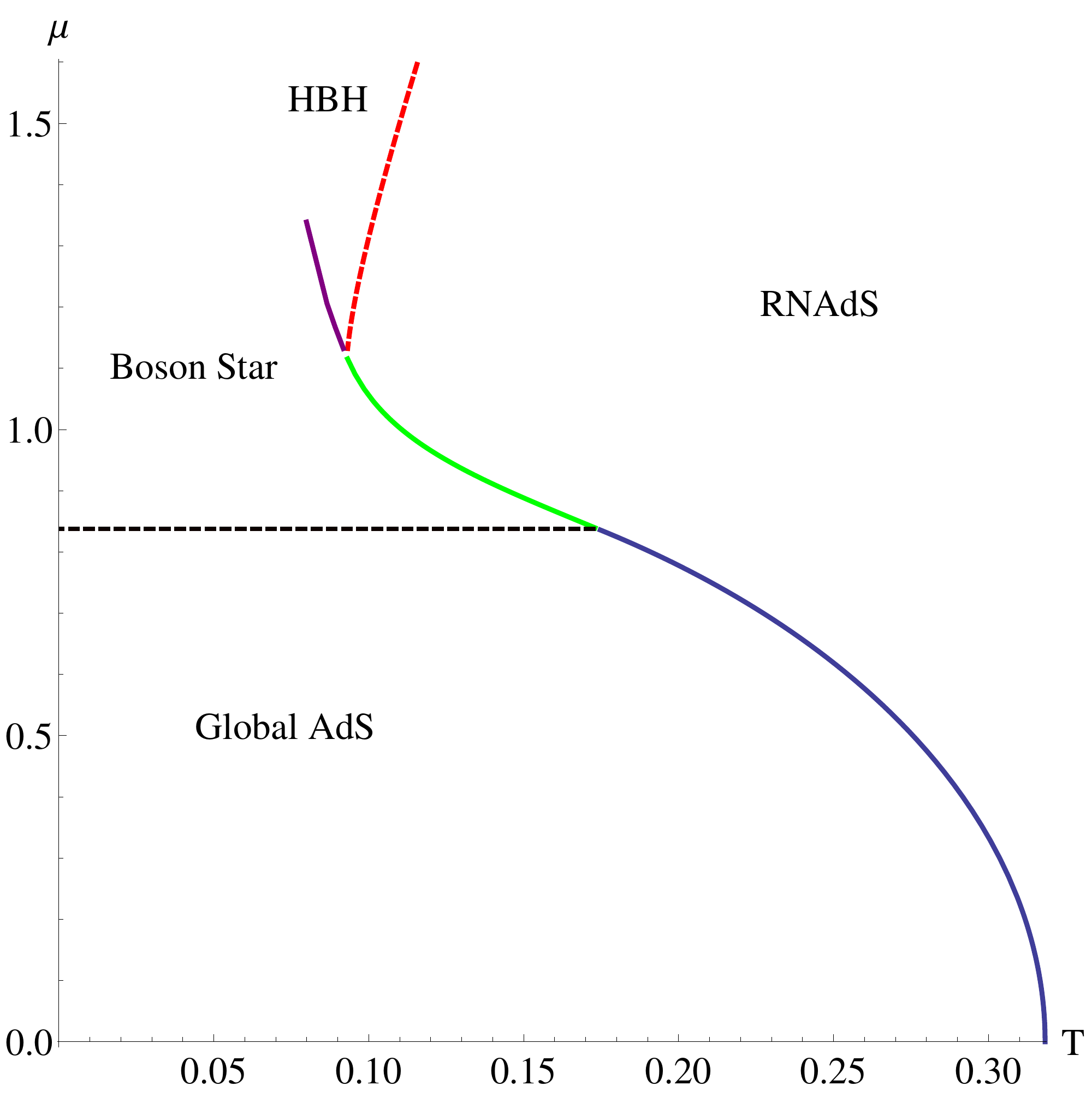}
        \caption{}
           \end{subfigure}
     \caption{Phase diagram calculated for $q=5,3,1.2$.}\label{q53}
\end{figure}

One thing to be noted is that for large $\mu$, the black hole is very big and the line of hair-forming instability asymptotes to a linear curve $\mu=\alpha T$, where the parameter $\alpha$ is $\frac{\mu}{T}$ in the Poincar\'e patch-AdS, and is dependent on $q$. 

The exact phase diagrams evaluated for $q=5,3,1.2$, are given in Fig.\ref{q53}.

\begin{figure}[H]
\begin{center}
\begin{tikzpicture}
\coordinate (a) at (0,0) {};
\coordinate (b) at (8,0) {};
\coordinate (c) at (0,8) {};
\coordinate (d) at (0,4.5) {};
\coordinate (e) at (4.5,0) {};
\coordinate (f) at (0,5) {};
\coordinate (g) at (4,7){};

\draw[ultra thick] (a) -- (b);
\draw[ultra thick] (a) -- (c);
\draw[blue,ultra thick] (e) to [out=90,in=0] node[midway, below right] {\;\;F-1} (d);

\draw[red,ultra thick,dashed] (f) to [out=0,in=235] node[midway, below right] {S-2} (g);
\node at (7.5,-0.5) {$T$};
\node at (-0.5,7.5) {$\mu$};
\node at (6,2.5) {$\framebox{RNAdS}$};
\node at (1.5,1) {$\framebox{global AdS}$};
\node at (1.2,6) {$\framebox{Hairy BH}$};

\end{tikzpicture}
\end{center}
\caption{Schematic phase diagram for $q<1$.}
\label{fig:phase2}
\end{figure}
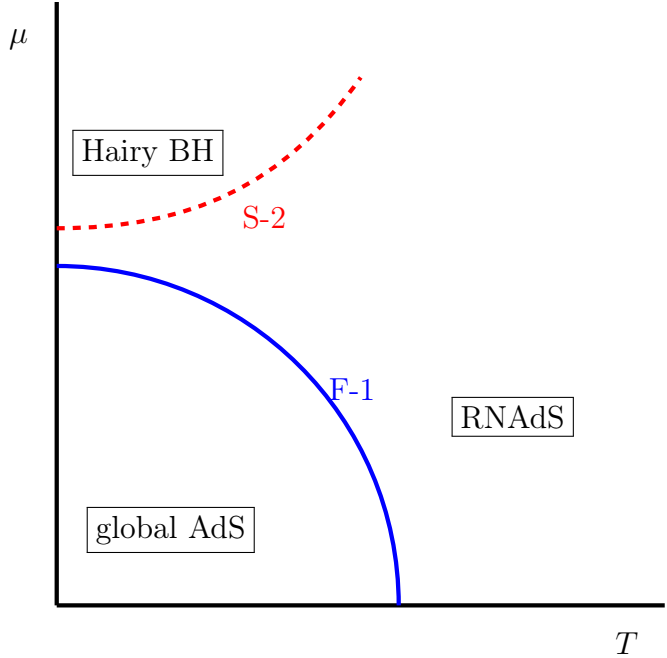

\subsection*{$ q \le 1 $}

\begin{figure}
    \centering
    \begin{subfigure}[b]{0.45\textwidth}
        \includegraphics[width=\textwidth]{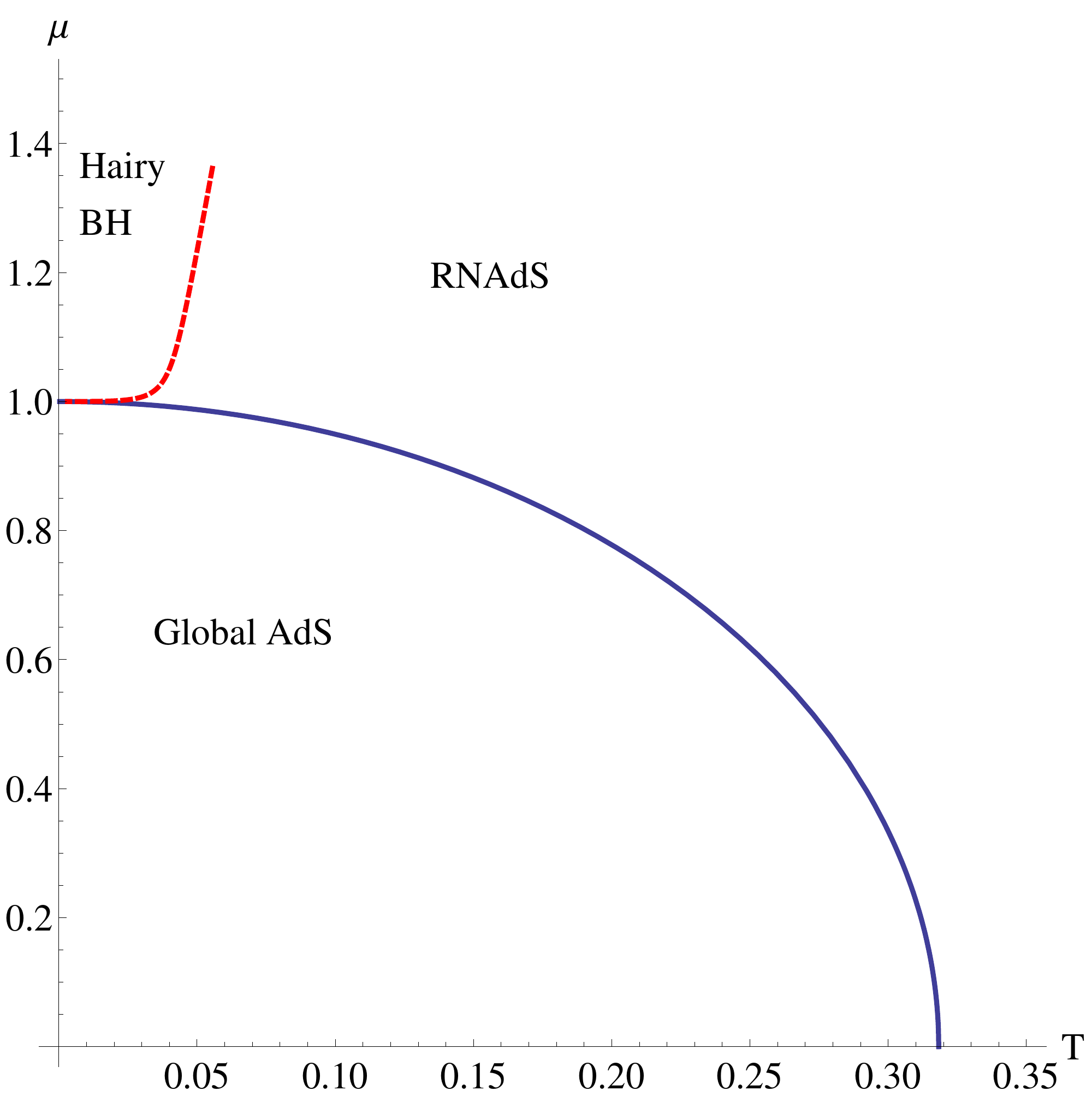}
        \caption{}
        \label{q1}
        \end{subfigure}
    \begin{subfigure}[b]{0.45\textwidth}
        \includegraphics[width=\textwidth]{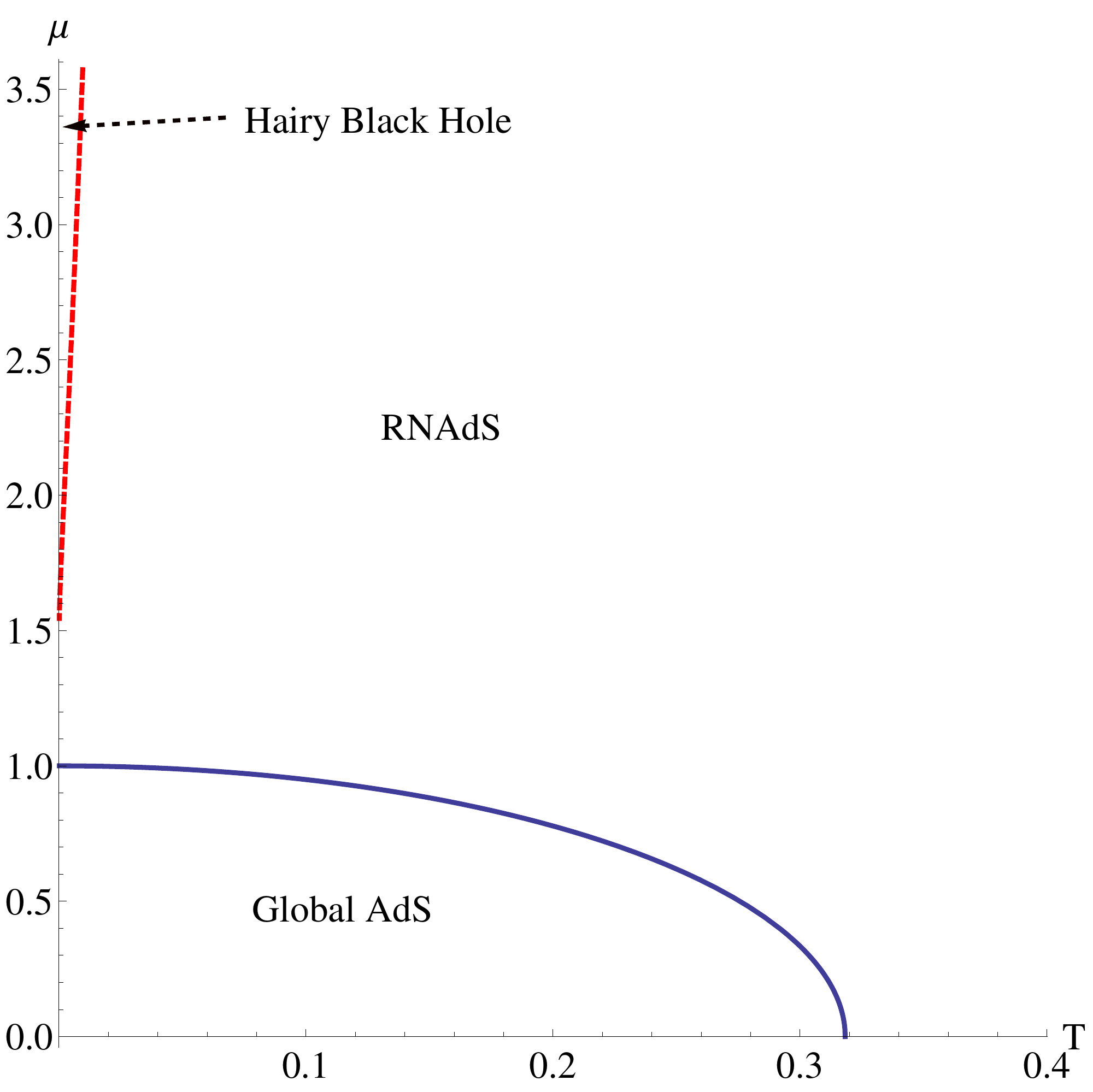}
        \caption{}
          \end{subfigure}
     \caption{Phase diagram calculated for $q=1,\frac{1}{2}$.}\label{ql1}
\end{figure}

At $q=1$, it can easily be seen from our analytic results that boson star instability happens at $\mu=1$. As the extremal RN black hole near $\mu=1$ has a radius approaching zero, the geometry is almost identical to the global AdS. Hence it is no surprise that line of hairy black hole instability and boson star instability coincides for $T=0,\mu=1$ and $q=1$ in Fig.\ref{q1}. 
 
We now look at $q<1$, where, as we had discussed earlier, the boson star instability happens at $\mu>1$. The transition from global AdS to boson star is a second order transition. So as we increase the chemical potential to $\mu>1$, the system undergoes a first order transition to RNAdS, the Hawking-Page transition (the curve \emph{F-1} in Fig.\ref{fig:phase2}), before the boson star instability could set in. Since the system is now in the black hole phase, the possible second order transition is the one in which the RNAdS develops a hair. This transition can be found as in the earlier case, numerically. The starting point of the instability curve (named here also as \emph{S-2}) is at the $T=0$ axis. 

The phase diagram in this case has only three different phases. The exact phase diagrams for $q=1,1/2$ are given in Fig.\ref{ql1}. In the case of $q=1$, the boson star and hairy black hole instability seem to happen at $\mu =1$. Here too, the number of phases remain three.

A Chandrasekhar-like instability for AdS black holes was found in \cite{Min1, Min2} when the $q$ was below some bound. In our case, that bound corresponds to $q=1$. We have checked that the for values of $q$ that are less than 1, the curvatures diverge at $r=0$ as we increase the condensate of the boson star. (But note that in our case, in this regime the boson star is not the dominant phase anyway.)

It is interesting to note that in our case we form a hairy block hole for arbitrarily small values of $q$ (see also, \cite{Sumit}). In the Appendix B, we discuss the extremal case analytically.

\subsection{Comments on Condensate Plots}

\begin{figure}
\centering
\includegraphics[width=0.5\textwidth]{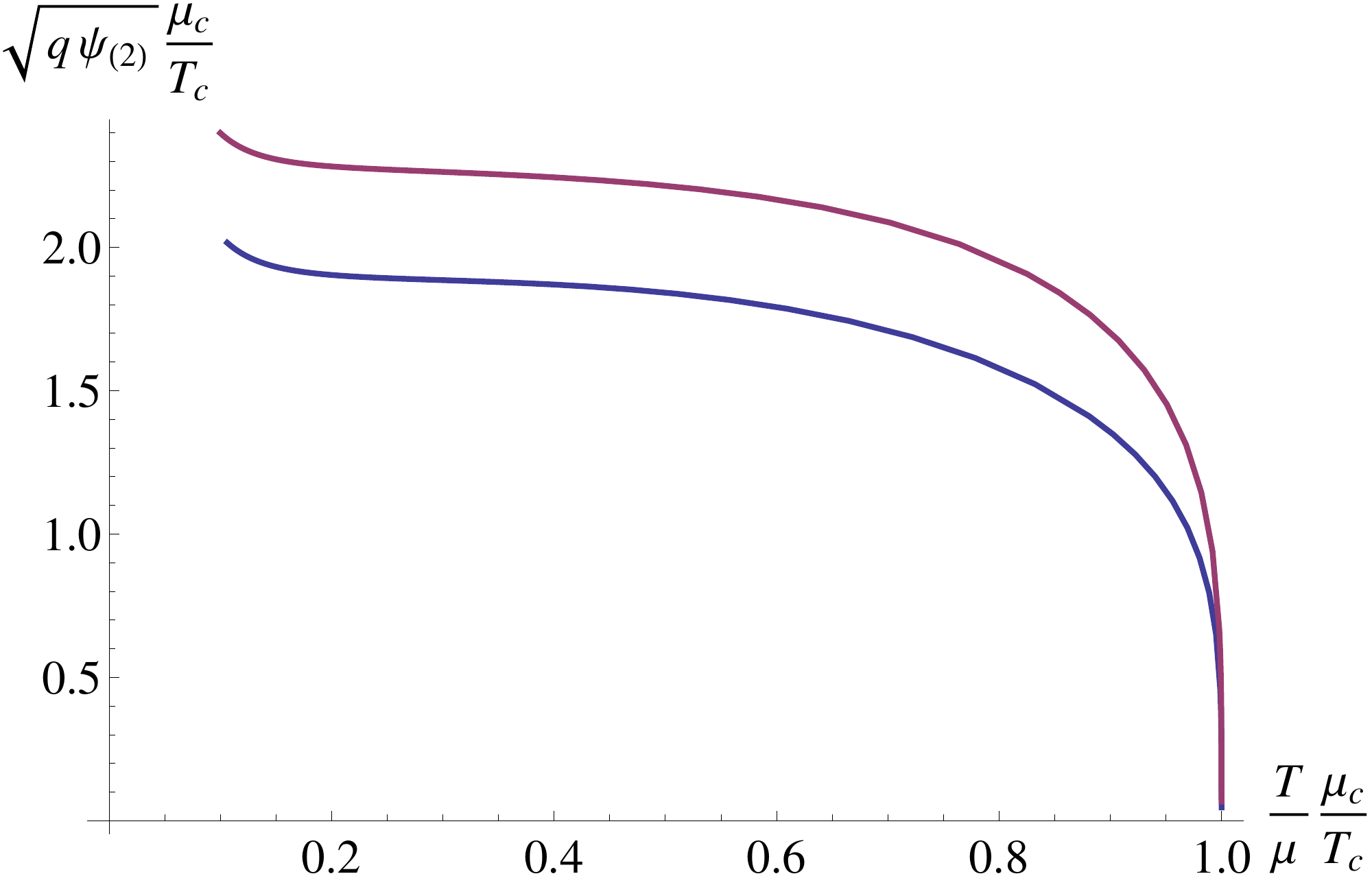}
\caption{Condensate plots for $q=10$ and $\rh = \tan^{-1}(0.9)$  [purple], $\tan^{-1}(1)$ [blue].}
\label{condensate}
\end{figure}

In the case of the Poincar\'e patch, the condensate plots were made for different values of $q$ alone, as there were two coordinate rescalings available. Here, however, one could plot for various values of $q$ and for different values of $\rh$. We have given the plot for $q=10$, for $\rh = \tan^{-1}(0.9),\tan^{-1}(1)$ in Fig.\ref{condensate}. Here $\mu_{c}$ and $T_{c}$ are defined to be the $\mu,T$ for the smallest value of $\psi_{0}$ considered -- this is the onset of the phase transition line (numerically in our computation it is $\approx 10^{-5}$). It can be seen that for larger $\rh$ the condensate decreases and the curve profile is slightly smaller. 
It should be noted that because we have two dimensionful quantities available, we have some freedom in choosing how to plot the condensate plots. 

\section{Concluding Comments}

In this paper we have investigated the phases of black holes in global AdS spacetime for large and moderate values of the scalar charge $q$. We have not investigated the phase structure for very small values of $q$ but we expect that it should not change qualitatively from the behavior we found for $q < 1$. It will be interesting to see if this expectation is correct, by doing a more exhaustive scan of the low charge regime. 

The instabilities/phases that we have uncovered are to be understood in the context of phases of thermal partition functions in the dual gauge theory. Recently there has been a lot of interest in various questions regarding non-linear dynamical instability of classical gravity in AdS \cite{instability}, which should be understood in the context of thermalization in the gauge theory. It will be interesting to try and fit these two perspectives into a coherent whole.  Another obvious line of development is to consider, in analogy with the work on Poncar\'e patch superfluids in \cite{superfluid}, adding a spatial component to the gauge field to the configurations considered here. 

\appendix
\section{Free Energy computation by background subtraction}
The free energy of a system can be computed by evaluating the classical action directly. However, the classical energy computed this way is divergent, and has to be made finite by using counter-terms or by using background subtraction, where we subtract out the classical action of the global AdS solution after matching the temperatures of the two configurations at the asymptotic region. 

Let us take the equations of motion to be of the form $R_{\mu\nu} - \frac{1}{2}g_{\mu\nu} R = \tilde{T}_{\mu\nu}$. It can be easily checked for our ansatz that
\bea
\mathcal{L} - R = \dfrac{2}{r^{2}}\tilde{T}_{33} = \dfrac{2}{r^{2}\sin^{2}\theta}\tilde{T}_{44}.
\eea

Now using $G_{\mu\nu}= R_{\mu\nu} - \dfrac{1}{2}g_{\mu\nu} R$, and $G^{\mu}_{\mu} = - R $, the classical action will simplify to the following form, in the Euclidean signature
\bea
S_{Euc} = -  \dfrac{1}{16\pi} 2\;{\rm Vol}_{3}\left(\int_{r_{0}}^{\infty}dr \sqrt{h(r)} - r g(r) \sqrt{h(r)}\Biggr|_{r\rightarrow\infty} \right),
\eea
where ${\rm Vol}_{3}= 4\pi \beta$. Also, $r_{0}=0$ for any solution without a horizon, namely global AdS and boson star, and $r_{0} = \rh $ for the RNAdS black hole and the hairy black hole solutions. The action calculated this way will be divergent and we can use holographic renormalization or background subtraction to regularize the action. We will be using background subtraction using global AdS to regularize and we get
\bea
S_{reg} = -  \dfrac{1}{2} \beta \left(\int_{r_{0}}^{\infty}dr \sqrt{h(r)} - r g(r) \sqrt{h(r)}\Biggr|_{r\rightarrow\infty} \right) + \beta_{0}\dfrac{1}{2}\left(\rb - \rb \;(1+\rb^{2}) \right)\Biggr|_{\rb\rightarrow\infty},
\eea
where $\beta$ and $\beta_{0}$ are the periodicities of the $t$ integrals of the geometry that we are interested in and of global AdS, respectively. Now, we have to adjust $\beta_{0}$ such that the geometry at the hypersurface $r=\rb $ of both the spaces are the same. This gives the relation
\bea
\dfrac{\beta}{\beta_{0}} &= &\sqrt{\dfrac{g_{tt}^{gAdS}(\rb)}{g_{tt}(\rb)}} = \sqrt{\dfrac{1+\rb^{2}}{g_{tt}(\rb)}},\\
\Rightarrow F &= &\dfrac{S_{reg}}{\beta} = -  \dfrac{1}{2} \left(\int_{r_{0}}^{\infty}dr \sqrt{h(r)} - r g(r) \sqrt{h(r)}\Biggr|_{r\rightarrow\infty} \right) + \dfrac{1}{2}\sqrt{\dfrac{g_{tt}(\rb)}{1+\rb^{2}}}\left(\rb - \rb \;(1+\rb^{2}) \right)\Biggr|_{\rb\rightarrow\infty}.
\eea

One can check that the classical action computed using this equation matches with that computed using \eqref{therm} analytically for RNAdS black hole, and numerically for the hairy solutions.

The evaluation of thermodynamic variables in the case of numerical hairy solutions is done by fitting the curves of $g(r), \phi(r)$ and we take $E$ to be $(-\frac{1}{2})$ times the coefficient of $\frac{1}{r}$ term of $g(r)$, and chemical potential($\mu $) and charge $(Q)$ from the falloff of $\phi(r)$, using the relation $\phi(r\rightarrow\infty) \approx 2 \mu - \frac{2 Q}{r}$.

\section{Extremal Black Hole instability}

In the full set of possible solutions, there are two systems with zero temperature and scalar condensate, boson star and extremal hairy black hole. 

The near-horizon instability of the extremal black hole is analytically tractable. The extremal RNAdS black hole has a charge, in terms of $\rh$ (with $L=1$),
\bea
Q = \pm 	2 \rh \sqrt{3 \rh^2 + 1}.
\eea
The chemical potential for the instability here is calculated as follows. We expand the scalar equation of motion in the extremal RNAdS background, in the near-horizon region. The leading order terms multiplying $\psi(r),\psi'(r),\psi''(r)$, together gives the equation of motion for a scalar in $AdS_{2}\times S^{2}$, which is the near-horizon geometry of $4$-d RNAdS. The instability point for the $4$-d extremal RNAdS is taken to be the value of $\mu$ that saturates the Breitolehner-Freedman bound for the $AdS_{2}$. This gives
\bea
-M_{(2)}^{2}L_{(2)}^{2}  =\dfrac{1}{4}=\dfrac{4q^2\mu^2 (\mu^2-1)}{3(2 \mu^2-1)^2}+\frac{2(\mu^2-1)}{3(2\mu^2-1)}.
\eea
We can solve for $\mu$ for a given value of $q$ for which the instability sets in and it is given by
\bea
\mu =  \sqrt{\frac{2 q^2}{4 q^2+1}+\frac{3}{2(4 q^2+1)}+\frac{ \sqrt{4 q^4+q^2+1}}{4 q^2+1}}.
\eea
The sign in the inside square-root has been fixed by noting that $\mu^2=5/2$ when $q=0$, which can be directly read off from the previous equation. For any value of $q$ (including $q=0$ !) this shows that for sufficiently large $\mu$ there is a near-horizon instability. 
 
We can use the above expression to compare the values of $\mu$ at which instability is triggered in the extremal and the boson star cases (the latter happens at $\mu =1/q$ as already noted). 
We can see that for large values of $q$, the boson star instability happens at a smaller value of $\mu$ than that of the extremal-RNAdS instability, and vice versa. We can also evaluate the $q$ at which the two curves intersect and it is found to be around $\approx 0.9$. 
We can think of the boson star instability as a proxy for the bulk instability of AdS black holes when the black hole is small (this is sometimes called superradiant instability, \cite{Min1, Min2}). It will be interesting to see  if the interplay between these two types of instabilities leads to a quantum critical phase transition for the hairy extremal solutions.


\end{document}